\documentclass[10pt,letterpaper]{article}

\pdfoutput=1

\usepackage{opex3_arxiv}

\usepackage{color}
\usepackage{graphicx}        
\usepackage{amsmath,amssymb}
\usepackage{dsfont}
\usepackage[isolatin]{inputenc}
\usepackage{caption}
\usepackage{subfig}

\begin{document}

\title{Intensity-only measurement of partially uncontrollable transmission matrix: demonstration with wave-field shaping in a microwave cavity}

\author{Philipp del Hougne$^1$, Boshra Rajaei$^{1,2}$, Laurent Daudet$^{1,3}$, Geoffroy Lerosey$^{1,*}$  }

\address{
$^1$Institut Langevin, ESPCI and CNRS UMR 7587, Paris, F-75005, France\\
$^2$Sadjad University of Technology, Mashhad, Iran\\
$^3$Paris Diderot University, Sorbonne Paris Cité, Paris, F-75013, France\\
}

\email{$^*$geoffroy.lerosey@espci.fr}

\begin{abstract}
Transmission matrices (TMs) have become a powerful and widely used tool to describe and control wave propagation in complex media. In certain scenarios the TM is partially uncontrollable, complicating its identification and use. In standard optical wavefront shaping experiments, uncontrollable reflections or additional sources may be the cause; in reverberating cavities, uncontrollable reflections off the walls have that effect. Here we employ phase retrieval techniques to identify such a partially uncontrollable system's TM solely based on random intensity-only reference measurements. We demonstrate the feasibility of our method by focusing both on a single target as well as on multiple targets in a microwave cavity, using a phase-binary Spatial-Microwave-Modulator.
\end{abstract}

\bibliographystyle{osajnl}


\section{Introduction}

In recent years, the control of wave propagation in complex media has received a lot of attention, being crucial to improve information transfer in a wide range of domains, including imaging, medical therapies and telecommunication amongst others \cite{InfoTransfer,NatPhotReview,park2013WSforOCT,choi2015WSforBioMed}. The fundamental problem of wave propagation in complex media is that the multiple scattering events result in a speckle as output wavefront that has lost any correlation with the input wavefront \cite{goodman_speckle}. Yet this wave diffusion process is linear and deterministic, which enabled the development of approaches to counteract the complete scrambling of the initial wavefront; notably time reversal and spatial wavefront shaping provide methods of identifying an input that then yields the desired output after propagation through the complex medium \cite{TR_fink,EMTR_prl,mosk_SLM}. It turned out that the multiple scattering, rather than being a problematic source of noise, provides additional degrees of freedom that can in fact be exploited to outperform focusing in homogeneous media \cite{EMTR_science,mosk_disorder4perfectFOC,choi_beatDifLim,park2013subwavelength}, or for sub-sampled compressive imaging \cite{liutkus14}.

In the context of focusing, be it with Spatial-Light-Modulators (SLMs) in the optical or Spatial-Microwave-Modulators (SMMs) \cite{SMM_design} in the microwave domain, the required input wavefront was initially identified with closed-loop iterative feedback schemes. These approaches are efficient to focus on a single point but have to be repeated if the target is changed. Subsequently, the more flexible tool of transmission matrices (TMs) has been explored \cite{popoff_prl}. Indeed, access to the TM of the system implies complete knowledge of the wave propagation inside the complex medium. Once identified, the TM enables direct focusing and image transmission at wish \cite{popoff_NatComm}, without any experimental trial-and-error, i.e. in open loop. Moreover, studies of the system's transmission eigenchannels to maximize energy transfer have been rendered possible by the TM \cite{choi_eigenchannels}. Notably in biomedical contexts, these approaches were successfully transferred to multimode fibers, which are a further example of a complex medium \cite{cizmar_biomed,choi_biomed,bianchi_biomed,pap_biomed}.

What all methods to identify the TM have in common is the need for reference measurements. In optics, as CCD sensors only provide intensity measurements, the TM identification started off with phase-shifting interferometric approaches \cite{popoff_prl}. It was later refined in terms of speed and technical requirements to increase its applicability in practical contexts \cite{park_largeTM}. Without resorting to such indirect phase measurements, signal processing techniques known as phase retrieval (PR) can also recover the full complex-valued TM only from the knowledge of the magnitudes of projections of known patterns through the medium. Applications of such PR techniques range from quantum mechanics \cite{corbett06} and nanoscience \cite{harrison93,PR_NANOscience,bunk07} to astronomical imaging \cite{fienup87}. More advanced PR algorithms, based on Bayesian modeling \cite{schniter15, rajaei16}, can explicitly take into account partial knowledge of signals and noise. In particular, such Bayesian-based PR algorithms allow the estimation of the TM with intensity-only measurements, corresponding to random inputs from binary-only amplitude modulation \cite{basis}.

In this study, we go a step further and consider systems whose TM is only partially controllable. The limited control over the TM might for instance be due to uncontrollable light sources or reflections that cannot be modulated by the SLM in optical experiments; it also occurs in the case of reverberating cavities. A practical example of the latter is found in the context of optimizing wireless communication in indoor environments (weakly reverberating cavities) by partially covering the walls with a SMM \cite{SMM_PoC}. By using intensity-only random reference measurements to feed the TM algorithm we keep technical requirements to a minimum, enabling the application of our method in situations where phase information might not be available or accessible. However, having some uncontrolled part in the TM brings new challenges, as this changes the statistics of the reference measurements. For focusing, this requires a specific global phase shift of the input wavefront - while the TM can only be estimated up to a global phase in each row. We develop the required theoretical framework based on the existing prSAMP phase retrieval algorithm \cite{rajaei16,rajaei15}. We then experimentally demonstrate the feasibility by focusing in a microwave cavity of low quality factor with a SMM that offers binary phase modulation, both for a single target and multiple targets. In addition to showing how signal processing techniques can be incorporated as a ``black-box'' in other areas of physics, this approach is particularly relevant for the previously evoked real-life telecommunication applications.

\section{Concept of partially uncontrollable TM}

\begin{figure}[h!!!]
\begin{center}
\includegraphics[width=\columnwidth]{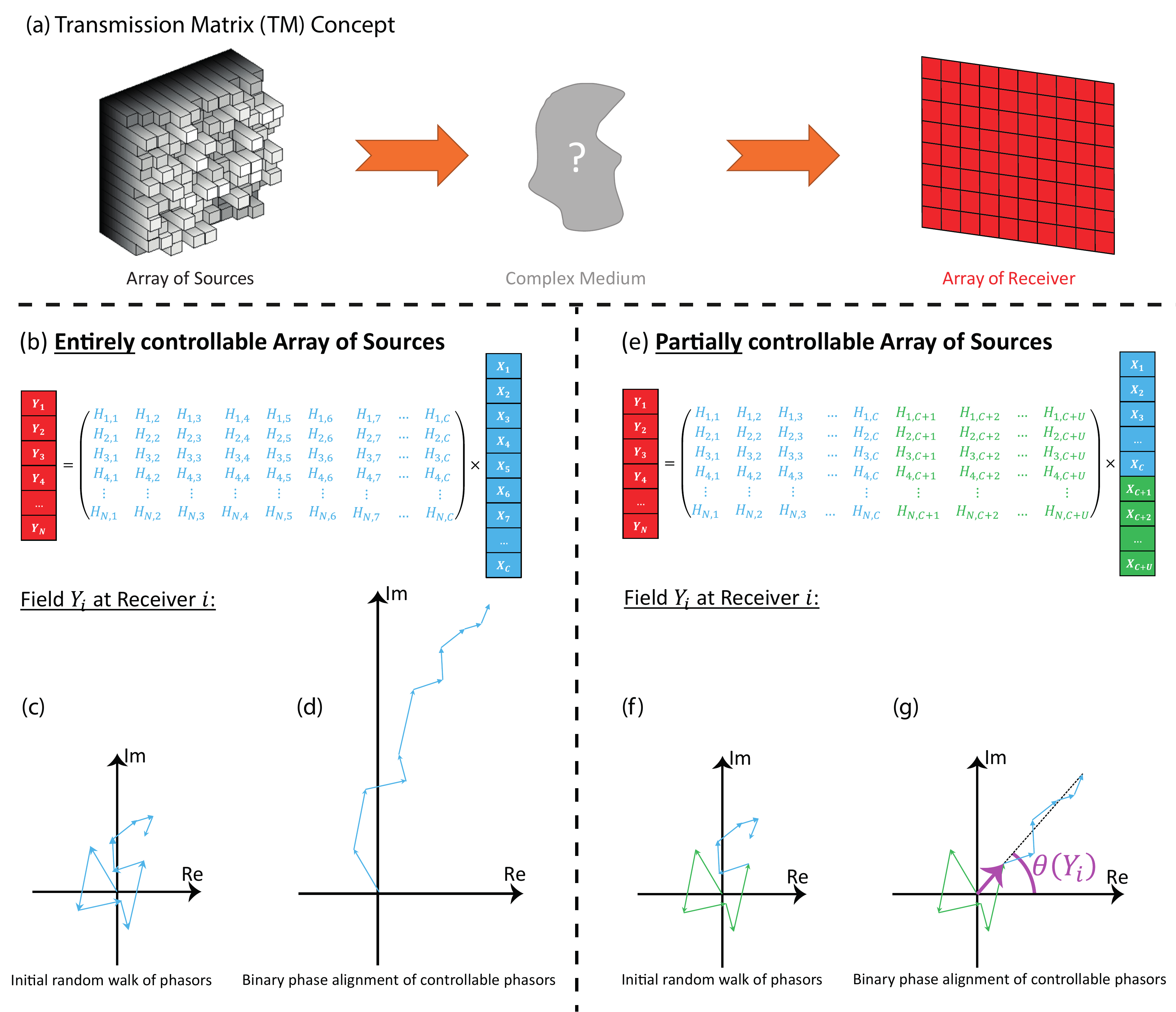}
\caption{Concept of wave focusing with a partially controllable TM: A wavefront originating from an array of sources with $C$ controllable and $U$ uncontrollable elements propagates linearly through a complex medium and is probed by an array of $N$ receivers. Initially each input mode has a random phase, such that their sum at one of the receivers constitutes a random walk in the Fresnel plane. To maximize the amplitude of the sum of all modes at one output position, (binary) phase alignment of the controllable modes is carried out. Note the importance of aligning the controllable phasors such that their sum has the same phase $\theta$ as the sum of all the uncontrollable modes in (g).}
\label{fig:Concept}
\end{center}
\end{figure}

A TM $\mathrm{\textbf{H}}$ relates the system's $N$ output modes $\mathrm{\textbf{Y}} = \{ Y_1, Y_2, ... , Y_N \}$ to its $M$ input modes $\mathrm{\textbf{X}} = \{ X_1, X_2, ... , X_M \}$:
\begin{align}
\mathrm{\textbf{Y}} = \mathrm{\textbf{HX}}.
\end{align}
As depicted in Fig. 1(a), the output modes are the signals picked up by a receiver array and the input modes correspond to the states of an array of sources; the latter can be an array of emitters (e.g. ultrasound transducers) or an array of reflectors (e.g. SLM, SMM) that act as secondary sources. The complex medium can be any medium causing wave diffusion and reverberation, such as a highly scattering material, a multimode fiber or a reverberating cavity, to name the most common examples.

The field at output point $i$ is the linear combination of the independent contributions from each of the sources:
\begin{align}
Y_i = \sum_{j} H_{i,j} X_j. 
\end{align}
Initially, each element of this sum can be considered as a random complex number that is conveniently represented as a phasor (arrow) in the complex Fresnel plane \cite{goodman_speckle}. The sum of all the random phasors can thus be visualized as a random walk (Fig. 1(c)). Focusing at position $i$ through phase modulation of the source array states, that is maximizing the received intensity $|Y_i|^2$, requires phase alignment of these phasors. If the entire array of sources is controllable (Fig. 1(b)), all of the phasors are controllable and the radial direction of alignment is irrelevant (Fig. 1(d)). This is the situation considered by standard TM schemes \cite{popoff_prl}. Binary phase control results in an imperfect zig-zag alignment, as shown in Fig. 1(d). To realize said phase alignment in practice, both closed-loop and open-loop schemes exist that are discussed in section 4.

Here we consider the case where $U$ source array elements are not controllable (Fig. 1(e)). In the context of the reverberating cavity of low quality factor the cavity modes have large line-widths such that there are more modes overlapping at the working frequency than can be controlled with the SMM \cite{publikation1}. This means that the phasors colored in green in Fig. 1(f) cannot be aligned. To maximize $|Y_i|^2$ nonetheless, the controllable (blue) phasors thus have to be aligned in a specific radial direction, namely that of the sum of the uncontrollable phasors; this angle $\theta(Y_i)$ is indicated in Fig. 1(g) and varies randomly with the output position $i$. This is the challenge we tackle in this paper, first for a single receiver and later for multiple receivers, with a TM approach.

\section{Experimental setup}

\begin{figure}[h!]
\begin{center}
\includegraphics[width=\columnwidth]{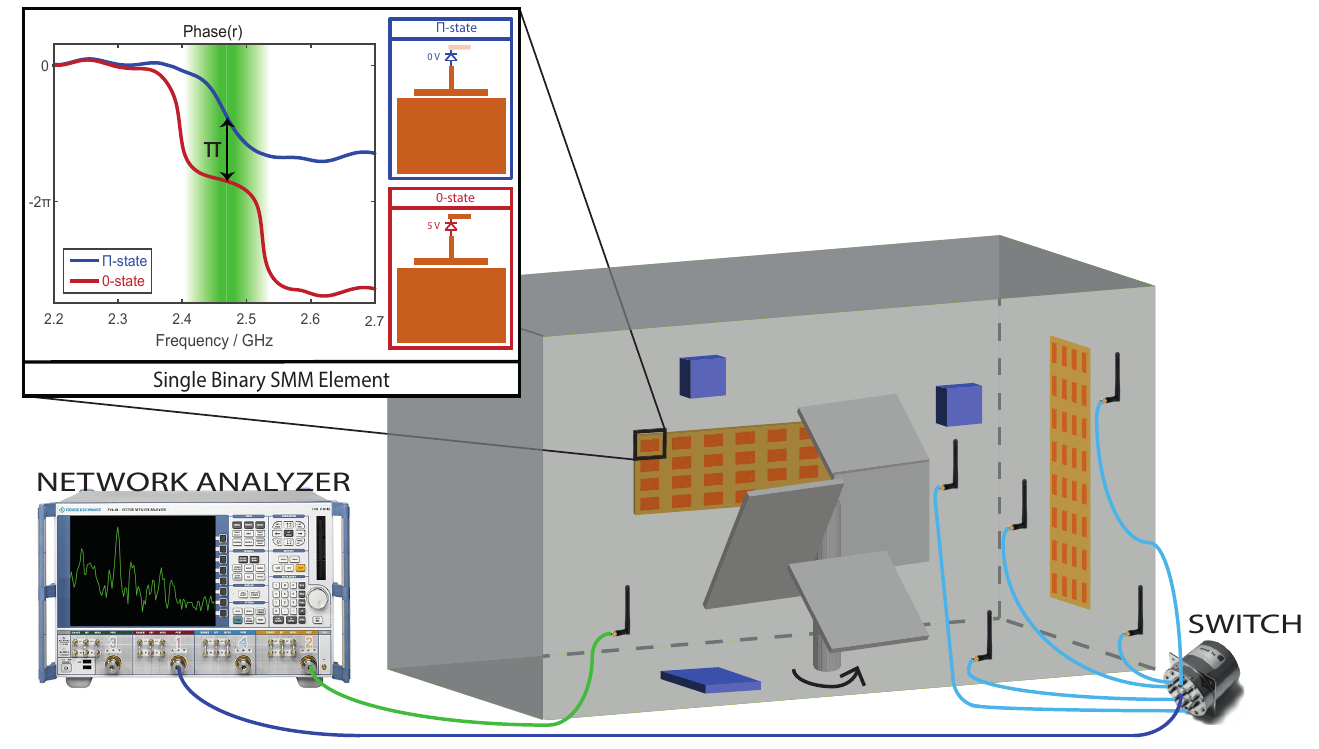}
\caption{Experimental scheme: A binary phase modulation SMM (see inset for details, adapted from~\cite{SMM_design}) partially covers the walls of a metallic cavity. Electromagnetic absorbers, isotropically distributed, reduce the cavity's quality factor. A mode-stirrer rotation of $12^\circ$ creates a statistically independent, ``new'' disordered cavity. This enables many realizations of disorder for averaging. We emphasize that only intensity information from the network analyzer is used.}
\label{fig:SetUp}
\end{center}
\end{figure}

Our experimental setup uses a binary phase modulation SMM whose working principle and characteristics are illustrated in the inset of Fig. 2 and in \cite{SMM_design} (working frequency $f_0 = 2.47 \ \mathrm{GHz}$). It partially covers the walls of a metallic cavity of dimensions $1.45 \ \mathrm{m} \times 1 \ \mathrm{m} \times 0.75 \ \mathrm{m}$. Electromagnetic absorbers are approximately isotropically distributed inside the cavity to ensure a low quality factor; this ensures that the SMM elements are independent and is representative of real-life telecommunication applications. A network analyzer is used to probe the transmission between one ``source'' antenna and a network of ``receiver'' antennas (in inverted commas to acknowledge spatial reciprocity). We emphasize that only intensity information from the network analyzer was used to prove the feasibility of our theoretical framework. As the outcome of a single realization in complex media focusing experiments is very dependent on the initial conditions, we average over many realizations of disorder. This is conveniently achieved by repeating the experiment after having rotated the mode-stirrer indicated in Fig. 2 by $12^\circ$; each rotation creates a statistically independent, ``new'' disordered cavity. Moreover, we average over three ``receiver'' positions.

\section{Single target focusing}

In this section we consider focusing (intensity maximization) on a single target, that is $\mathrm{\textbf{Y}}$ consists of a single receiver, to compare how the optimized input configurations we identify based on the TM perform in comparison to a closed-loop experimental feedback scheme. Firstly, we briefly revisit the iterative algorithm. Secondly, we introduce the theoretical framework employed to calculate the TM with the existing prSAMP phase retrieval algorithm \cite{rajaei16,rajaei15}. Thirdly, we present two methods to compute the required input for focusing, based on the identified TM. Finally, we evaluate the two TM-based approaches in comparison with the experimental closed-loop iterative scheme.

\subsection{The experimental closed-loop iterative scheme}

The task of the algorithm is to identify the input wavefront configuration that maximizes the transmitted intensity at the selected target position, by aligning all controllable phasors with the uncontrollable contribution. We use an iterative algorithm that tests for one SMM element after the other which of the two states yields the higher intensity at a target \cite{moskWSalgo}; this state is kept and the next element is tested, etc.
The enhancement $\eta_E$ is quantified by the ratio of the averaged target intensity after and before focusing: 
\begin{align}
\eta_E = \frac{\langle|Y|^2\rangle_{final}}{\langle|Y|^2\rangle_{initial}}. 
\end{align}

\begin{figure}[h!]
\begin{center}
\includegraphics[width=\columnwidth]{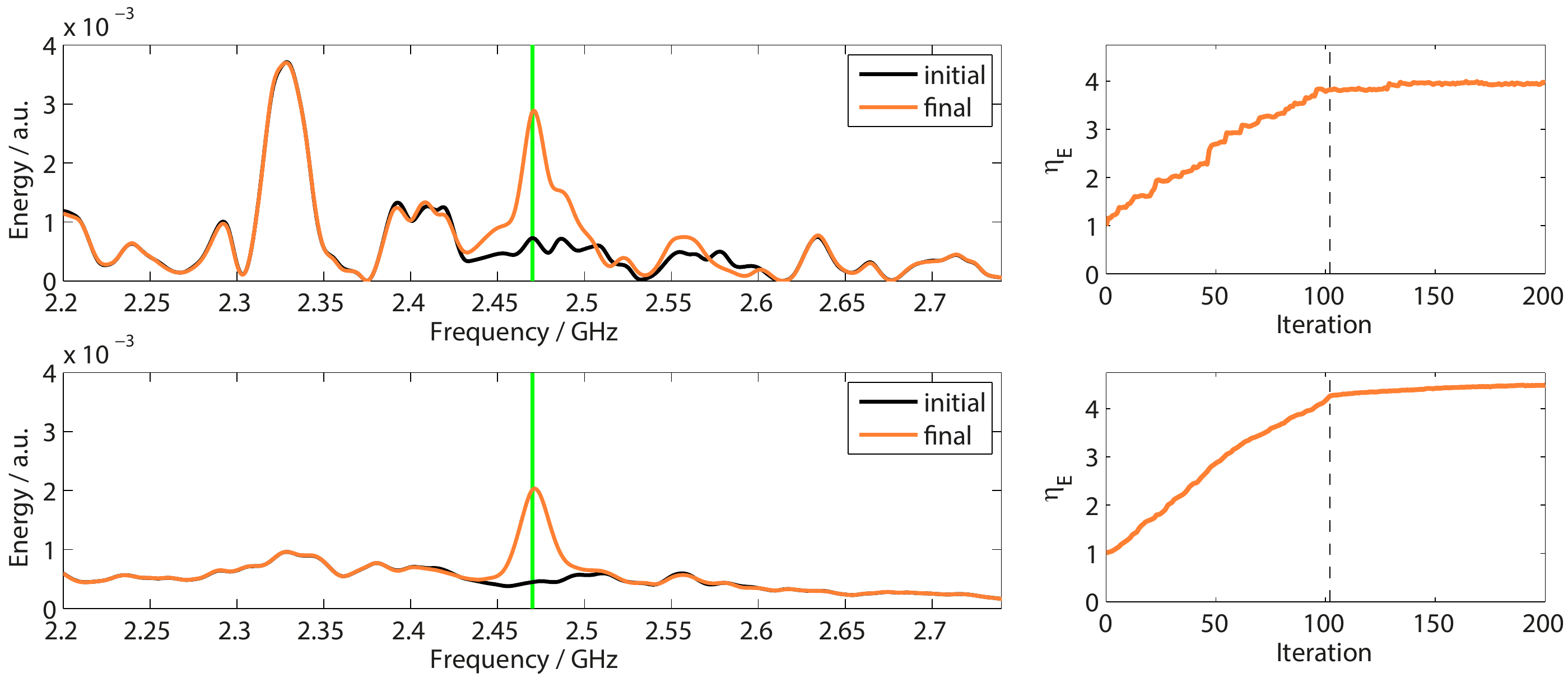}
\caption{Closed-loop iterative focusing on a single target: The spectra before (black) and after (orange) optimization are shown, both for a single realization (top row) and averaged over $90$ realizations (bottom row). Moreover, the corresponding optimization dynamics are displayed, that is how the focusing progresses with each iteration in terms of the target intensity enhancement $\eta_E$. The dashed line indicates the iteration after which all $102$ SMM pixels have been tested once.}
\label{fig:Iterative}
\end{center}
\end{figure}

In Fig. 3 the iterative algorithm's performance is displayed both for a single realization and averaged over disorder. The optimization dynamics show that after having tested each SMM element once, the enhancement $\eta_E$ saturates. This rapid convergence is only possible because the optimum state of a given element is independent of the other elements' states and because the uncontrollable contribution dominates such that even during the first iterations the correct phase alignment for the considered elements is identified. The former can be attributed to the low quality factor of our weakly reverberating cavity, the latter to the fact the only $6 \ \%$ of the cavity surface are covered by the SMM. In high quality factor cavities wave propagation is of course still linear but the strong reverberation causes correlations between different TM elements, complicating both identification and use of the TM. The spectra after optimization show a clear peak centered on the working frequency. The peak width of $26 \ \mathrm{MHz}$ corresponds to the cavity's correlation frequency \cite{SMM_PoC}; in general, it is the smaller out of modulator bandwidth (here $66 \ \mathrm{MHz}$, cf. inset of Fig. 2) and the medium's correlation frequency.

\subsection{Theoretical framework required to use existing phase retrieval algorithms}

Phase retrieval (PR) algorithms aim at recovering a complex-valued signal $\mathbf{u}\in \mathbb{C}^N$ from the magnitude of its (possibly noisy) projections $| \mathbf{M}\mathbf{u} |$, where $\mathbf{M} \in \mathbb{C}^{M\times N}$ is a known matrix called measurement matrix. PR has a number of applications in imaging, including X-ray crystallography \cite{harrison93}, X-ray diffraction imaging \cite{bunk07} and astronomical imaging \cite{fienup87}. Many algorithms have been reported to solve the PR problem, mostly in cases where $\mathbf{H}$ is the Fourier transform or a random matrix with random iid Gaussian coefficients. These methods include, but are not limited to, alternating projections such as Gerchberg and Saxton \cite{gerchberg72} and Fienup \cite{fienup78} and several variants of them \cite{marchesini07,netrapalli15}, convex relaxation algorithms such as phaseLift \cite{candes13} and phaseCut \cite{waldspurger15} and spectral recovery methods \cite{alexeev14}.

For the estimation of a TM with intensity measurements, we collect a number $P$ of measurements as intensity of the outputs, written in matrix form as $\mathbf{Y}=[\mathbf{y}_1,\ldots,\mathbf{y}_P]$, corresponding to the
source array inputs $\mathbf{X}=[\mathbf{x}_1,\ldots,\mathbf{x}_P]$. Conjugate-transposing the transmission equation $\mathbf{Y} = \mathbf{HX}$ gives
\begin{equation}
\mathbf{Y}^T =  \mathbf{X}^T \mathbf{H}^H.
\label{eq:transposeTM}
\end{equation}
Hence, a PR algorithm can be used to estimate the complex-valued TM $\mathbf{H}$, row-by-row, from intensity measurements $|\mathbf{Y}|$. It has to be noted that the TM can only be estimated up to a global phase on each row, due to the fundamental phase invariance of the measurements. While this is of no consequence when the entire field is controlled by the TM, this brings additional complexity in the general case of a partially controlled TM, as described below. Note also that using binary patterns as inputs of the reference measurements (rows of the measurement matrix $\mathbf{X}^T$) makes the PR problem ill-conditioned, such that most PR algorithms fail to recover $\mathbf{H}$. Hence, we here resort to a Bayesian-based PR algorithm called prSAMP (phase retrieval Swept Approximate Message Passing) that was precisely designed to handle such ill-conditioned measurement matrices, with possibly a high level of noise (the interested reader may refer to \cite{rajaei16} for a detailed discussion on the prSAMP algorithm).

In the current setup, due to the uncontrollable contribution of the field, we cannot simply feed the algorithm with our reference measurements $(\mathrm{\textbf{X}},|\mathrm{\textbf{Y}}|)$. The probability density function of $|\mathrm{\textbf{Y}}|$ is here of modified Ricean nature, rather than Gaussian \cite{goodman_speckle, publikation1, S12_statistics}. Instead, we split the right hand side of Eq. 1 into its controllable and uncontrollable contributions, indicated by $c$ and $u$, respectively.
\begin{align}
\mathrm{\textbf{Y}} = \mathrm{\textbf{H}}_c \mathrm{\textbf{X}}_c + \mathrm{\textbf{H}}_u \mathrm{\textbf{X}}_u.
\end{align}
We note that due to its very nature the uncontrollable part is fixed and we should use the prSAMP algorithm to identify $\mathrm{\textbf{H}}_c$, to then align the controllable modes with $ \mathrm{\textbf{H}}_u \mathrm{\textbf{X}}_u$. To enable the use of prSAMP as a ``black-box'' we have to remove the contribution of $ \mathrm{\textbf{H}}_u \mathrm{\textbf{X}}_u$ from $|\mathrm{\textbf{Y}}|$. Averaging over many random SMM configurations yields $\langle\mathrm{\textbf{X}}_c\rangle = \mathrm{\textbf{0}}$ and thus 
\begin{align}
\langle\mathrm{\textbf{Y}}\rangle = \langle\mathrm{\textbf{H}}_c \mathrm{\textbf{X}}_c + \mathrm{\textbf{H}}_u \mathrm{\textbf{X}}_u\rangle = \mathrm{\textbf{H}}_c \langle\mathrm{\textbf{X}}_c\rangle + \mathrm{\textbf{H}}_u \mathrm{\textbf{X}}_u = \mathrm{\textbf{H}}_u \mathrm{\textbf{X}}_u.
\end{align}
Given that random reference measurements are required anyway to use prSAMP, we can also use them to remove the uncontrollable contribution from $|\mathrm{\textbf{Y}}|$. We can then feed the prSAMP algorithm with $(\mathrm{\textbf{X}},||Y_i|-\langle|Y_i|\rangle|)$ which follows the assumed Gaussian distribution. Thereby we compute the controllable part of the TM row corresponding to output point $i$. Due to the lack of any phase information, the algorithm computes $\mathrm{\textbf{H}}_c$ up to a random global phase offset $e^{i\epsilon}$ on each row of the TM. In previous works, with entirely controllable TMs, this did not pose a problem for intensity maximization. In the present case, however, the aligned modes have to be in phase with the uncontrollable contribution (see Fig. 1). 

Once again we can make use of the available reference measurements, in combination with a simple phase retrieval based on alternating projections \cite{bauschke02}. The relative angle between two vectors can be identified if their individual amplitudes and the amplitude of their sum is known; with the knowledge of many realizations this can be done unambiguously. Suppose that the true phase angle of $\mathrm{\textbf{H}}_u \mathrm{\textbf{X}}_u$ is $\theta$, then we can identify the unique angle $(\theta+\epsilon)$.

\subsection{TM-based identification of optimal input wavefront}

The remaining question is that of how to ultimately identify the input $\mathrm{\textbf{X}}_{c-opt}$ that yields optimal focusing; as discussed previously, this requires both maximization of $|\mathrm{\textbf{H}}_c \mathrm{\textbf{X}}_c|$ and phase alignment with $\mathrm{\textbf{H}}_u \mathrm{\textbf{X}}_u$. Firstly, based on the matrix formalism from Eq. 1, we consider an inversion approach that can be applied to maximize $|\mathrm{\textbf{H}}_c \mathrm{\textbf{X}}_c|$; given our experimental constraint of binary-only phase modulation, it turns out to be a simple binary phase conjugation:
\begin{align}
\mathrm{\textbf{X}}_{c-opt} = \mathrm{sign}[\mathrm{Re}(\mathrm{\textbf{H}}_c^{-1}\mathrm{\textbf{Y}}_{obj})] = \mathrm{sign}[\mathrm{Re}(\mathrm{\textbf{H}}_c^{-1})].
\end{align}
Taking the sign of the real part ensures that $\mathrm{\textbf{X}}_{c-opt}$ is binary and hence $\mathrm{\textbf{Y}}_{obj} = 1$ can be used. Of course $|\mathrm{\textbf{H}}_c \mathrm{\textbf{X}}_{c-opt}|$ is not aligned with $|\mathrm{\textbf{H}}_u \mathrm{\textbf{X}}_u|$ yet. 
There is a difference of $\epsilon$ between the actual phase angle of $\mathrm{\textbf{H}}_c\mathrm{\textbf{X}}_{c-opt}$ and the ``desired'' one of $\mathrm{phase}(\mathrm{\textbf{Y}}_{obj})=\mathrm{phase}(1)=0$. In a second step a rotation by $(\theta+\epsilon)$ ensures alignment with the uncontrolled part, without explicit knowledge of $\epsilon$.

Secondly, we propose a hybrid iterative-TM approach that does not rely on a TM inversion. We rotate $\mathrm{\textbf{H}}_c$ (a single row for a single target) by the phase of $\mathrm{\textbf{H}}_u \mathrm{\textbf{X}}_u$ and then starting from a random initial configuration iteratively optimize $\mathrm{\textbf{X}}_c$, with the feedback calculated using the TM. This is similar to the closed-loop iterative algorithm used in section 4.1 but we emphasize that here only numerical calculations using the estimated TM rather than experimental measurements are performed to provide the feedback. The computation time of the simple matrix products involved is well below a second with any computer. To be sure that the global optimum is identified the procedure can be repeated several times although the iterative algorithm in this context tends to almost always identify the global maximum in one go.

\subsection{Experimental evaluation}

We compare the performance of the experimental closed-loop iterative scheme and the two TM-based schemes for single target focusing, in dependence of the number of reference inputs $K$ available for the phase retrieval algorithm. For each value of $K$, we repeat the experiment for three independent ``receiver'' positions and thirty independent mode-stirrer positions, thereby carrying out $90$ statistically independent realizations of the experiment.

\begin{figure}[h!]
\begin{center}
\includegraphics[width=\columnwidth]{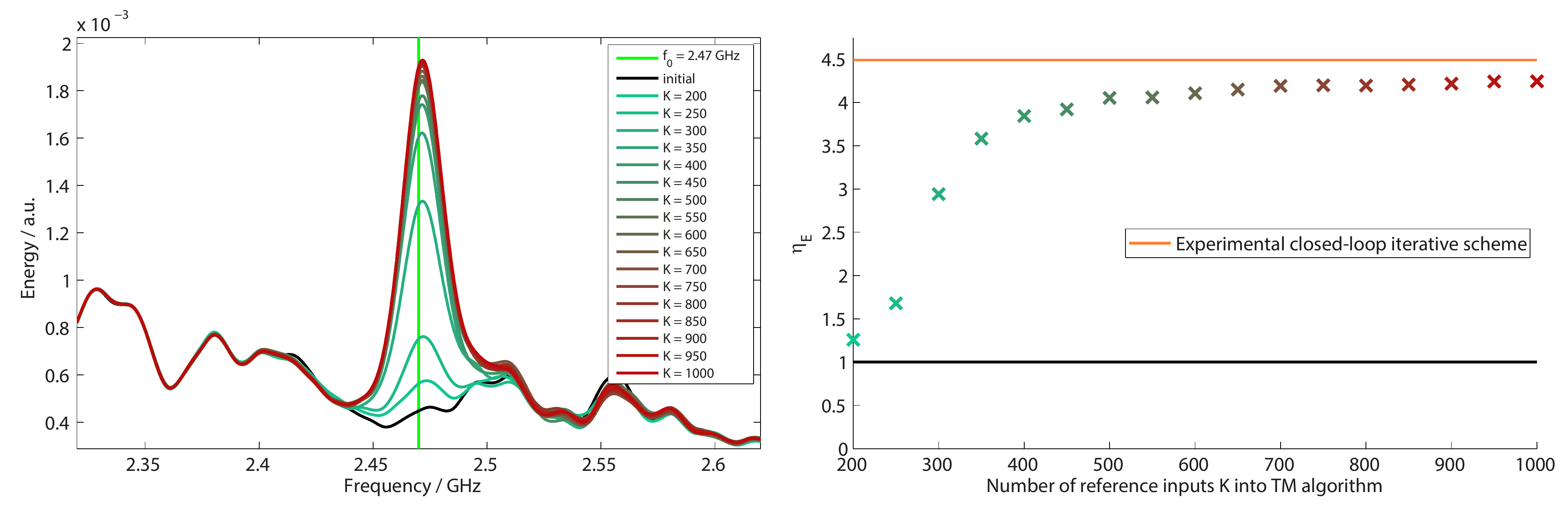}
\caption{Comparison of TM-based identification of optimum input configuration for focusing with the experimental closed-loop iterative scheme (orange), as a function of the number of available reference inputs $K$ for the phase retrieval algorithm that computes the TM. Note that both TM-based approaches yielded identical results. The left hand side shows the averaged spectra corresponding to the focusing results based on $K$ reference inputs.}
\label{fig:SingleTarget}
\end{center}
\end{figure}

Fig. 4 displays the averaged results. Firstly, note that both TM-based approaches yielded identical results. Secondly, notice that it only takes about $K=400$ reference inputs for the algorithm to produce reasonable results that are only slightly inferior to the closed-loop iterative scheme ($86 \ \%$). For $K>400$, the attainable enhancement then slowly saturates. Even though the TM-based results remain slightly below the achievable outcome of the iterative scheme ($95 \ \%$), Fig. 4 proves our ability to compute the TM even when it is partially uncontrollable and to use it for focusing on a single target. Obviously, single target focusing only constitutes a proof of concept rather than being an end in itself: at least $400$ reference measurements are four times more than the number of iterations required by the closed-loop scheme to saturate, as seen in Fig. 3. It is in the context of multiple targets that the TM becomes powerful, as discussed in the next section.

\section{Multiple targets focusing}

In this section we consider the challenge of simultaneously focusing on multiple targets based on the estimated TM. As the phase retrieval algorithm computes the TM row-by-row (that is target-by-target), each row of $\mathrm{\textbf{H}}_c$ ends up having its own global phase error, which means that the TM inversion based method from before cannot be applied successfully. We employ thus only the hybrid approach in this section that we have shown to work equally well in the previous section. A trade-off seems to naturally arise between maximizing the intensity averaged over all targets and ensuring reasonable equality amongst all targets (to avoid essentially focusing on only some of the targets). An important detail to evoke is hence what quantity we actually optimize in the case of multiple targets; here we minimize the mean distance between desired amplitude $A$ and actual amplitude $|Y_i|$, raised to a power $p$, averaged over all targets:
\begin{align}
D = \langle| A - |Y_i| |^p \rangle_{targets}.
\end{align}
Intuitively, the higher $p$ is, the more emphasis should be given to the equality amongst the targets. In principle different values of $A$ can be chosen for each target to enable the transmission of complicated patterns or images through the complex medium. It is worth noting that minimizing $D$ in the special case of a single target is directly equivalent to maximizing the target intensity, as done in the previous section.

In the case of multiple targets, we define $\eta_E$ as before but additionally averaged over the targets. Fig. 5 presents the averaged spectra corresponding to focusing on up to $n_{targets} = 10$ targets, with $p=2$. (We found in practice that the impact of $p$ on $\eta_E$ was close to negligible, $\eta_E$ decreased very slightly for higher values of $p$. The equality amongst the targets improved for higher values of $p$ but was reasonable for all values. For clarity we hence limit ourselves to the case of $p=2$ here.)

\begin{figure}[h!]
\begin{center}
\includegraphics[width=\columnwidth]{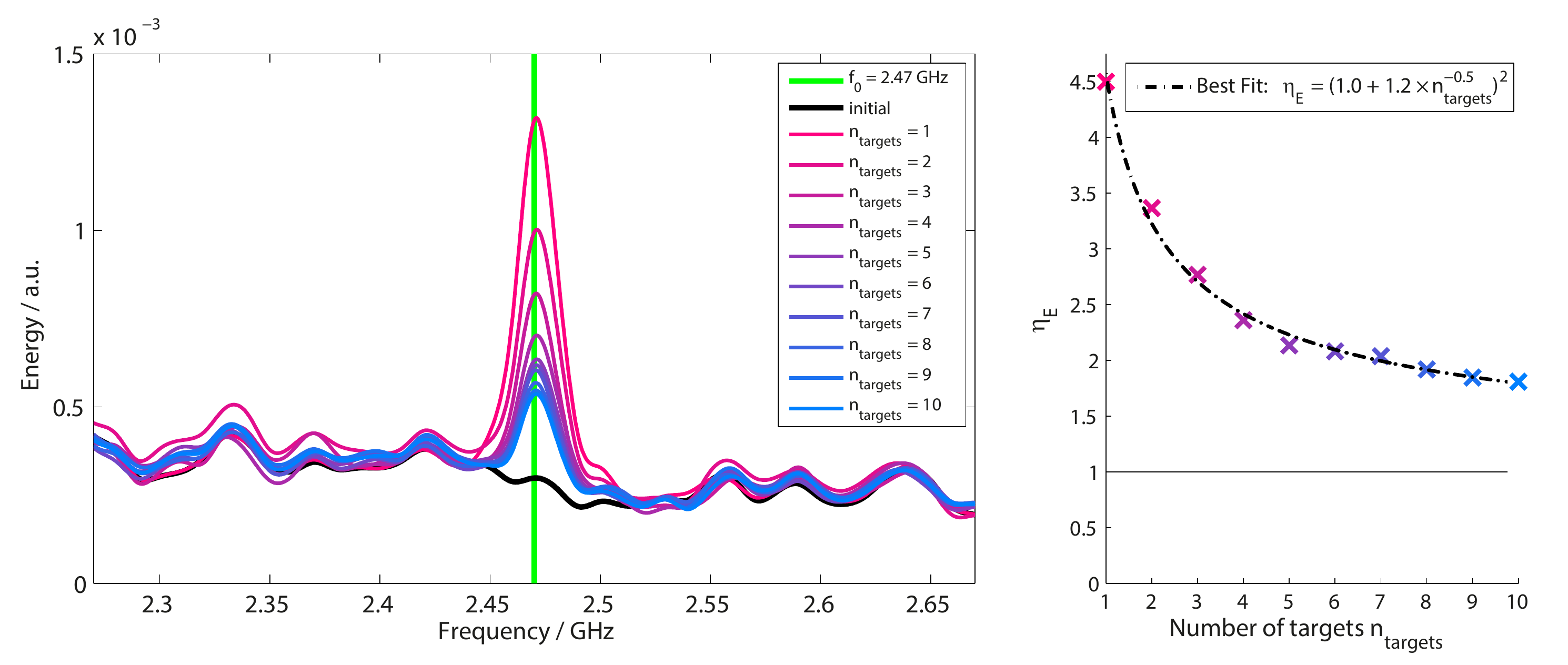}
\caption{TM-based focusing on multiple targets ($K=800$). The spectra averaged over realizations of disorder and the $n_{targets}$ targets after focusing on the left, and the corresponding enhancement averaged over all targets on the right. The best fit with the corresponding equation is indicated. }
\label{fig:mpf}
\end{center}
\end{figure}

These results clearly demonstrate that despite the limitation of our SMM to phase-binary modulation and despite the uncontrollable contribution, TM-based focusing on multiple targets is doable. The average focusing intensity decreases as $1/n_{targets}$ because there is essentially a certain amount of energy that the SMM can relocate and that is distributed equally across all targets on average. In terms of the intensity enhancement rather than the final intensity this takes the following form:
\begin{align}
\eta_E (n_{targets}) = \Big(\eta_A (n_{targets})\Big)^2 = \bigg(1 + \frac{\eta_A(1)-1}{\sqrt{n_{targets}}}\bigg)^2,
\end{align}
where $\eta_A(n_{targets}) = \Big\langle\frac{\langle|Y|\rangle_{final}}{\langle|Y|\rangle_{initial}}\Big\rangle_{targets}$ is the average amplitude enhancement. Eq. 9 provides an excellent fit in Fig. 5. The fact that in a weakly reverberating cavity we still managed to almost double the transmitted energy even when focusing on ten targets is very relevant to telecommunication applications whose parameters correspond directly to our experiment.

\section{Conclusion}

This paper shows that even in complex media with a partially uncontrollable transmission matrix, the \textit{complex-valued} transmission matrix can be estimated, up to a global phase factor on each of its rows, purely based on \textit{intensity-only} output measurements corresponding to random, phase-binary input patterns. In our experiment, the inputs are phase modulated with a binary Spatial-Microwave-Modulator that partially covers the walls of a weakly reverberating cavity. The global phase offsets on each TM row are not a limiting factor as we resort to a hybrid TM-iterative approach rather than TM inversion: the TM is used to calculate the feedback for the iterative algorithm. Focusing on multiple points is demonstrated with an approach that in principle also enables the transmission of complicated patterns and images through a complex medium with partially uncontrollable TM.

The presented methods widen the range of TM applications, from the obvious implementation in telecommunication within weakly reverberating office rooms to optical set-ups that suffer from uncontrollable light sources. This work is an example of how signal processing techniques, here phase retrieval, can be employed in a ``black-box'' fashion in other areas of physics, with real-life applications.

\section*{Acknowledgments}

P.d.H. acknowledges funding from the French ``Ministère de la Défense, Direction Générale de l\textquotesingle Armement''. B.R. acknowledges funding from PSL under grant CSI:PSL. This work has been supported by LABEX WIFI (Laboratory of Excellence within the French Program ``Investments for the Future'') under references ANR-10-LABX-24 and ANR-10-IDEX-0001-02 PSL* and by Agence Nationale de la Recherche under reference ANR-13-JS09-0001-01.

\end{document}